\documentclass[prd,12pt,aps,prd,longbibliography,nofootinbib]{article}

\usepackage{amsmath,amssymb,graphicx,multirow,xspace,slashed}
\usepackage[colorlinks=true,urlcolor=blue,anchorcolor=blue,citecolor=blue,filecolor=blue,linkcolor=blue,menucolor=blue,pagecolor=blue]{hyperref}
\usepackage[compress,numbers]{natbib}
\usepackage{subfigure, placeins}
\usepackage{booktabs}
\usepackage{blindtext, rotating}
\usepackage{afterpage}
\usepackage{enumitem}
\usepackage{ marvosym }
\usepackage{authblk} 
\usepackage{verbatim}
\usepackage{soul} 
\usepackage[normalem]{ulem}
\usepackage{pifont}

\usepackage{floatrow}
\newfloatcommand{capbtabbox}{table}[][\FBwidth]

\usepackage[font=footnotesize,labelfont=bf]{caption}

\usepackage{lineno}

\allowdisplaybreaks

\long\def\symbolfootnote[#1]#2{\begingroup%
\def\thefootnote{\fnsymbol{footnote}}\footnote[#1]{#2}\endgroup}

\newcommand{\newc}{\newcommand}
\newc{\gsim}{\lower.7ex\hbox{$\;\stackrel{\textstyle>}{\sim}\;$}}
\newc{\lsim}{\lower.7ex\hbox{$\;\stackrel{\textstyle<}{\sim}\;$}}
\newc{\gev}{\,{\rm GeV}}
\newc{\mev}{\,{\rm MeV}}
\newc{\ev}{\,{\rm eV}}
\newc{\kev}{\,{\rm keV}}
\newc{\tev}{\,{\rm TeV}}
\newc{\MHT}{$H_T^{\text{miss}}$}
\newc{\MET}{$\slashed{E}_T$}
\newc{\MTT}{$M_{T2}$}

\newc{\mz}{M_Z}
\newc{\mpl}{M_*}
\newc{\mw}{m_{\rm weak}}
\newc{\nr}[1]{N^c_R{}_{#1}}

\def\beq{\begin{equation}}
\def\eeq{\end{equation}}
\newcommand{\bea}{\begin{eqnarray}\begin{aligned}}
\newcommand{\eea}{\end{aligned}\end{eqnarray}}
\def\bitem{\begin{itemize}}
\def\eitem{\end{itemize}}
\newcommand{\CO}{O}

 \numberwithin{equation}{section}

\newcommand\fverb{\setbox\fverbbox=\hbox\bgroup\verb}

\newbox\fverbbox

\usepackage[usenames,dvipsnames]{xcolor}

\definecolor{darkgreen}{rgb}{0,0.5,0}
\definecolor{goodorange}{rgb}{0.9,0.4,0}

\begin{document}

\baselineskip 0.6cm

\begin{titlepage}

\thispagestyle{empty}

\begin{center}

\vskip 1cm

{\Large \bf Pulling Out All the Tops with}\vskip 0.3cm {\Large\bf Computer Vision and Deep Learning}

\vskip0.2cm{}

\vskip 1.0cm
{\large Sebastian Macaluso and David Shih }
\vskip 1.0cm
{\it NHETC, Dept.~of Physics and Astronomy\\ Rutgers, The State University of NJ \\ Piscataway, NJ 08854 USA} \\
\vskip 2.0cm

\end{center}

\begin{abstract}

We apply computer vision with deep learning -- in the form of a  convolutional neural network (CNN) -- to build a highly effective boosted top tagger. Previous work (the ``DeepTop" tagger of Kasieczka et al) has shown that a CNN-based top tagger can achieve comparable performance to state-of-the-art conventional top taggers based on high-level inputs. Here,  we introduce a number of improvements to the DeepTop tagger, including architecture, training, image preprocessing, sample size and color pixels. Our final CNN top tagger outperforms BDTs based on high-level inputs by a factor of $\sim 2$--3 or more  in background rejection, over a wide range of tagging efficiencies and fiducial jet selections. 
 As reference points,  we achieve a QCD background rejection factor of 500 (60) at 50\% top tagging efficiency for fully-merged (non-merged) top jets with $p_T$ in the 800--900~GeV (350--450~GeV) range. 
Our CNN can also be straightforwardly extended to the classification of other types of jets, and the lessons learned here may be useful to others designing their own deep NNs for LHC applications.

\end{abstract}

\end{titlepage}

\setcounter{page}{1}

\tableofcontents

\vfill\eject

\section{Introduction}\label{Introduction}

Heavy boosted particles play an important role in many analyses at the LHC, including SM precision measurements, Higgs and electroweak physics, and searches for physics beyond the Standard Model (BSM). In general, the collimated decay products of boosted particles are reconstructed as a single large-radius ``fat jet". Analyses then attempt to ``tag" the origin of the fat jet by looking at its substructure. (For reviews of boosted object tagging and jet substructure, and many original references, see e.g.~\cite{Salam:2009jx,Abdesselam:2010pt,Altheimer:2012mn,Shelton:2013an,Altheimer:2013yza, Adams:2015hiv,Cacciari:2015jwa,Larkoski:2017jix}.) The ability to accurately tag boosted jets has many benefits. For instance, it can be used to overcome the QCD background and measure $h\to bb$ in associated production \cite{Butterworth:2008iy}. In BSM physics,  new heavy particles could be created, which then produce boosted SM objects as they decay.  Requiring the presence of these boosted objects is then a useful handle in discriminating signal against SM background. 

In this paper, we will focus on a particularly well-motivated case: boosted top jets. Signatures with energetic top quarks are predicted from SM processes such as single top and top pair production, and in several models of new physics. Top partners are expected to play a key role in solutions to the hierarchy problem, and they can naturally produce boosted top quarks in their decays. Additionally, there are other models that consider the production of dark matter in association with a top quark or top quark pair. Some recent LHC searches based on boosted top jets include \cite{Aaboud:2018eqg, Sirunyan:2018gka, ATLAS-CONF-2017-037, Sirunyan:2017wif}.

Traditional top tagging methods (see \cite{Plehn:2011tg,Kasieczka:2018ohx} for reviews and original references) start with a collection of physical observables, such as jet mass, that can be used to distinguish tops from light-flavor QCD. These high-level features can  serve as inputs to various multivariate machine learning algorithms, such as boosted decision trees (BDTs), to further enhance the tagger performance. These algorithms attempt to find  a set of complicated boundaries over the phase space that maximizes the classification accuracy. However, as the classification ability is highly dependent on these observables, the main challenge resides in finding ways to systematically come up with a set of observables that are not highly correlated and give the best discriminating power. 

By contrast, in recent years, there has been a great deal of interest in using deep neural networks (NNs) to identify objects at the LHC (among many other potential applications). The tantalizing promise of deep learning is the ability to start from much lower level inputs than was previously thought possible, and transform them into meaningful outputs. (For pedagogical introductions to neural networks and deep learning, see e.g.~\cite{Nielsen,Goodfellow-et-al-2016}.) In the context of boosted jet tagging, the idea is to allow a NN to figure out on its own, from relatively raw data (e.g.\ momentum four-vectors of all the constituents of the jet), the intrinsic patterns that identify each type of jet and the regions of phase space that distinguish them.  In this sense, deep learning attempts to {\it invent} the most useful physical observables for classification, in addition to designing the optimal set of cuts on these observables. 

The interest of the LHC community in deep learning has been spurred by the huge successes of deep NNs in real-world applications (see \cite{DeepLearning} for a nice overview). One major source of breakthroughs has been in computer vision,  from pixel level labeling of images for autonomous vehicles \cite{ROB:ROB20276, DBLP:journals/corr/abs-1202-2160} and Google's automatic captioning of images \cite{DBLP:journals/corr/VinyalsTBE14, 10.1007/978-3-642-15561-1_2}, to Facebook's DeepFace project \cite{6909616} and Microsoft surpassing human-level performance on ImageNet classification \cite{DBLP:journals/corr/HeZR015}. These results were made possible in large part thanks to the invention of convolutional neural networks (CNNs).
CNNs are built from two types of layers: \textit{convolutional layers} and  \textit{fully connected layers}. The former implement locality (exploit the image's spatially local correlations) and capture the lower level features of the input image (lines, edges, curves, etc.). These are eventually passed on to the latter which are responsible for learning abstract, higher level concepts (such as class probabilities). This independence from hand engineered features is a major advantage of CNNs from more traditional algorithms.

CNNs have a direct application to classifying jets at the LHC, since there is an obvious and natural sense in which jets can  be viewed as images. Indeed the calorimeters already provide the requisite pixelization. The intensity of the image can be the per-pixel $p_T$ and can be augmented with other per-pixel quantities such as track multiplicity. This idea of jet images has been explored in a number of works
\cite{Cogan:2014oua, Almeida:2015jua, deOliveira:2015xxd,Baldi:2016fql, Komiske:2016rsd,  Kasieczka:2017nvn}, with \cite{deOliveira:2015xxd,Komiske:2016rsd,Kasieczka:2017nvn} applying CNNs to $W$-boson, quark/gluon and top tagging respectively. These works have
demonstrated that jet taggers based on computer vision can perform comparably to or slightly better than conventional taggers based on high-level inputs.
In particular, the CNN top tagger of  \cite{Kasieczka:2017nvn} (named ``DeepTop" there) was trained on grayscale images formed from calorimeter deposits of moderately boosted top jets. The end result was a CNN top tagger with  performance comparable to state-of-the-art BDTs built out of \textsc{SoftDrop} variables \cite{Larkoski:2014wba}, HEPTopTaggerV2 (HTTV2) variables  \cite{Plehn:2009rk, Plehn:2010st, Kasieczka:2015jma}, and N-subjettiness \cite{Thaler:2010tr}. 

In this paper, we explore  a number of improvements to the DeepTop tagger, including the NN architecture (augmenting the DeepTop CNN with more feature maps and more nodes on dense layers), the NN training (loss function, optimizer algorithm, minibatch size, learning rate), image preprocessing, sample size (increasing the training sample by $10\times$ to $\sim$~1M jets saturates the NN performance), and adding color (calorimeter $p_T$, track $p_T$, track multiplicity and muon multiplicity). The result is a much more effective CNN for top tagging, one that (for the first time) significantly outperforms best-in-use conventional methods. 
This shows the enormous power and promise of modern deep learning methods as applied to the LHC. We are clearly entering a new era driven by major gains in artificial intelligence.

In order to disentangle any possible correlations between our proposed improvements and the fiducial jet image selection, we consider two very different jet samples in this paper.\footnote{We thank Gregor Kasieczka for  very stimulating discussions on this point.} The first is the sample of moderately-boosted jets used in the DeepTop paper ($350~{\rm GeV}<p_T<450~{\rm GeV}$). The second is a sample of high $p_T$ jets ($800~{\rm GeV}<p_T<900~{\rm GeV}$) that 
(apart from some minor differences) is taken from a recent note on top tagging methods by CMS \cite{CMS:2016tvk}. We will refer to these as the ``DeepTop sample" and the ``CMS sample" throughout this work.  Apart from the $p_T$ ranges, an important difference between the two samples is the {\it merge requirement}. This is a generator-level cut that requires the daughters of the top quark to be contained within the cone of the fat jet.  It  ensures that all the top jets contain the bulk of the energy from the boosted top quark. Without the merge requirement, the top jet sample is significantly polluted by  partially merged top jets that might contain only the $W$-boson, or only the $b$ quark and a single jet from the $W$ decay. The CMS sample imposes a merge requirement, while the DeepTop sample does not, and we will see that this has a major impact on the tagger performance.

Combining all of our proposed improvements, we show that the net effect is to increase the background rejection rate of the DeepTop tagger by a factor of $\sim3$--10 in the CMS sample, and a factor of $\sim 1.5$--2.5 in the DeepTop sample. It is perhaps not surprising that the improvement is much more modest in the DeepTop sample, since this was the focus of \cite{Kasieczka:2017nvn}. In any event, our modifications result in significant gains in performance over the baseline tagger for both jet samples, which is strong evidence for their general applicability. In both cases, the single greatest improvement is actually in the NN training, then followed by the NN architecture and the larger training sample size. This illustrates that the performance of a NN can be determined as much by the methods used to train it and the dataset it is trained on, as it is by the architecture. 

We then proceed to a comparison of our CNN top tagger with conventional top taggers that are meant to represent the state-of-the-art and best-in-use. For the DeepTop sample, we compare directly against the ``MotherOfTaggers" BDT ROC curve in fig.~8 of \cite{Kasieczka:2017nvn}. For the CMS sample, we compare against a BDT built out of HTTV2 variables and N-subjettiness. A cut-based version of this tagger was shown in  \cite{CMS:2016tvk} to have optimal performance among cut-based taggers (see also the analogous ATLAS references \cite{ATLAS-CONF-2017-064,ATL-PHYS-PUB-2017-004}). The upshot is that our CNN top tagger outperforms these conventional taggers  by a factor of  $\sim 2$--3 or more in background rejection, across a wide range of tagging efficiencies.

Very recently there have been several efforts  \cite{Louppe:2017ipp, Pearkes:2017hku, Butter:2017cot,Egan:2017ojy}  
 to feed the raw jet constituents (as momentum four-vectors) to various deep learning architectures such as recurrent neural networks (RNNs) and dense neural networks (DNNs). These have shown much promise. In \cite{Louppe:2017ipp} they showed that a recurrent neural network (RNN) $W/Z$ tagger can outperform a simple cut-based classifier based on N-subjettiness and jet mass.  In \cite{Pearkes:2017hku,Egan:2017ojy} they showed that a dense neural network (DNN) and an RNN top tagger can significantly outperform a likelihood-based tagger that takes N-subjettiness and jet mass as inputs. It would be extremely interesting to do a head-to-head comparison of all of these deep learning NNs with each other and with a state-of-the-art conventional tagger.

Although we have focused on top quarks in this work, it can also be viewed as a case study of boosted object tagging more generally. Our approach could be straightforwardly extended to other types of jets.  There are also many other potential applications (many have already begun to be explored), for instance whole-event classification, event generation, weakly-supervised learning, pile-up mitigation to name a few.
Furthermore, our optimizations were not systematic due to computational limitations. So perhaps with a more systematic approach (i.e.\ hyperparameter scans) one could achieve even greater gains. 

This paper is organized as follows. In section \ref{Methodology} we describe the details of our simulations and the precise specifications of our top and QCD jet image samples. We also briefly review the original DeepTop CNN which forms the baseline for the CNN tagger developed in this work, as well as the conventional taggers that we benchmark against. In section \ref{NN improvements} we give an overview of some general ``best practices" in the design of NNs, and we show how these can be applied to improve the DeepTop CNN. We hope that, apart from the usefulness of the CNN top tagger itself, this overview of concepts in NN design will prove useful to others. While much or all of it will be known to experts, it may be useful to have it collected in one place.
 
In Section \ref{Preprocessing}, we describe improvements to the image preprocessing steps in the DeepTop paper that are made possible by using the higher-resolution tracks in the jet. In Section \ref{Other improvements}, we examine the dependence of the classification accuracy on the training sample size and multiple intensity channels (colors). Then, in  Section \ref{Final comparison} we put it all together and compare our top tagger against the DeepTop tagger  and the conventional taggers built out of high-level inputs. We conclude with a discussion of next steps and promising future directions in Section \ref{Outlook}.
In Appendices \ref{Validation DeepTop} and \ref{Validation HEPTopTaggerV2} we validate our implementation of the DeepTop paper, and the cut-based CMS top tagger (using the HEPTopTaggerV2 and $\tau_{32}$ variables) respectively. In Appendix \ref{Merge requirement} we discuss the differences in top tagger performance if fully-merged-tops are required or not.

\section{Methodology}\label{Methodology}

The fat jets used in this paper are taken from all-hadronic $t \bar{t}$ and QCD dijet events generated in  proton-proton collisions  using \textsc{Pythia 8.219} \cite{Sjostrand:2014zea}, where multiparton interactions and pileup are turned off for simplicity. After showering and hadronization, the events are passed into  \textsc{Delphes~3.4.1} \cite{deFavereau:2013fsa} for detector simulation.  The jets are clustered with \textsc{FastJet~3.0.1} \cite{Cacciari:2011ma}.

As discussed in the Introduction, we will study improvements to the DeepTop tagger using two very different samples of jet images. These are described in table \ref{tab:jetsamples}. The first is the jet sample used in the DeepTop paper \cite{Kasieczka:2017nvn}, while the second is essentially the same as the high $p_T$ sample used in the CMS note  \cite{CMS:2016tvk}.\footnote{CMS uses $800~{\rm GeV}<p_T<1000~{\rm GeV}$ jets with $R=0.8$.} Let's now highlight some of the important differences between the samples: 

\begin{itemize}

\item The DeepTop sample is much lower $p_T$ than the CMS sample.

\item The DeepTop sample uses only calorimeter energies, while the CMS sample uses particle-flow, meaning that the tracks and neutrals (defined to be calorimeter towers minus the track contributions) are counted separately. This is very advantangeous, as the tracks have much higher resolution than the calorimeter towers.

\item With the  tracking information in the CMS sample, we can use color images along the lines of \cite{Komiske:2016rsd}. In addition to the colors used in \cite{Komiske:2016rsd} (calorimeter $p_T$ of the neutrals, per-pixel track $p_T$, and per-pixel track multiplicity), we also include muon multiplicity. This is motivated by the presence of muons in a sizable fraction of top quark jets coming from semileptonic $b$ decays. (For comments on $b$-tagging see Section \ref{Outlook}.) 

\item The DeepTop sample used  a toy calorimeter with resolution $\Delta\eta=0.1$, $\Delta\phi=5^\circ$. For the CMS sample we used the default CMS detector card that comes with \textsc{Delphes~3.4.1}, which has a slightly higher calorimeter resolution. The number of pixels ($37\times 37$) chosen for the high $p_T$ jet images is based on this. In both cases, a large image size is chosen to make absolutely sure the entire fat jet is captured.

\item 
Finally, a crucial difference between the two samples is the merge requirement. DeepTop did not require the daughters of the top quark to fall in the cone of the fat jet, while CMS did. With the merge requirement, the top jets are more ``top-like" (otherwise they are significantly contaminated by $W$ jets and $b$ jets), and this increases the potential discriminating power against QCD jets. Accordingly, we will see that the ROC curves for the CMS sample look much better than for the DeepTop sample. We explore this further in Appendix \ref{Merge requirement}. 
\end{itemize}

\begin{table*}[t]
\begin{tabular}{|c|c|c|}
\hline
& DeepTop  & CMS \\ \hline
\multirow{4}{*}{Jet sample} & 14 TeV & 13 TeV \\ & $p_T\in (350,450)$~GeV, $|\eta|<1$ & $p_T\in (800,900)$~GeV, $|\eta|<1$ \\ & $R=1.5$ anti-$k_T$ & $R=1$ anti-$k_T$\\ & calo-only & particle-flow \\& match: $\Delta R(t,j)<1.2$ & match: $\Delta R(t,j)<0.6$ \\ & merge: NONE & merge: $\Delta R(t,q)<0.6$
\\\hline
\multirow{2}{*}{Image} & $40\times 40$ & $37\times 37$ \\ &$\Delta \eta=4$, $\Delta \phi={10\over9}\pi$ & $\Delta \eta=\Delta\phi=3.2$
\\\hline
Colors & $p_T^{calo}$ &  $(p_T^{neutral},  \, p_T^{track},\, N_{track},\, N_{muon})$
\\\hline
\end{tabular}
\caption{The two jet image samples used in this work.
\label{tab:jetsamples}}
\end{table*}

We will benchmark our CNN top tagger against BDT taggers built out of high-level inputs. For the DeepTop sample, we directly compare against their ``MotherOfTaggers" BDT that takes HTTV2 variables, SoftDropped masses, and N-subjettiness variables (with and without SoftDrop) as inputs. Since we have fully validated the DeepTop minimal tagger, we do not bother to validate the MotherOfTaggers BDT as well, but just take its ROC curve directly from fig.~8 of the DeepTop paper. For the CMS sample, we will consider both a cut-based tagger that combines the HTTV2 variables with the N-subjettiness variable $\tau_3/\tau_2$ (motivated by the recent CMS note on top tagging \cite{CMS:2016tvk}), as well as a BDT trained on these variables. For the former, we varied simple window cuts on each of the variables, as in \cite{CMS:2016tvk}. We validate our implementation of this by reproducing the ROC curve shown in fig.~7R of \cite{CMS:2016tvk} using our own simulations (see appendix \ref{Validation HEPTopTaggerV2} for details). For our BDT we used the ROOT package TMVA \cite{Hocker:2007ht} with the same hyperparameters as in  \cite{ATL-PHYS-PUB-2017-004} and trained on the same jets as our final CNN tagger. 

For the design of  our CNN, we took as a starting point the DeepTop tagger of   \cite{Kasieczka:2017nvn}.  Its CNN architecture consisted of four identical convolutional layers (8 feature maps, $4\times 4$ kernel) separated in half by one $2\times 2$ max-pooling layer, followed by three fully connected layers  of 64 neurons each and an output layer of two softmax neurons. Zero-padding was included before each convolutional layer to prevent spurious boundary effects. The DeepTop CNN was trained on a total of 150k+150k top and QCD jet images, by minimizing a mean-squared-error loss function using the stochastic gradient descent algorithm in minibatches of 1000 jet images and a learning rate of 0.003. In order to validate our implementation of the DeepTop tagger, we have carefully reproduced the ROC curve in fig.~8 of  \cite{Kasieczka:2017nvn}, see appendix \ref{Validation DeepTop}  for details. 

Using the DeepTop tagger, the authors of   \cite{Kasieczka:2017nvn} demonstrated that CNNs could perform comparably to a conventional BDT trained on high-level inputs. 
In the following sections we will consider a number of improvements to the DeepTop tagger that, taken together, demonstrate for the first time that CNNs can significantly outperform conventional taggers.

\section{Improvements to the neural network}\label{NN improvements}

In the design of an effective neural network, there are countless choices to be made. These include not only decisions about the neural network architecture (how many layers, of what type), but also how it is trained (loss function, optimizer, minibatch size, etc). In general, the many parameters that go into the design of a neural network are referred to as ``hyperparameters" (not to be confused with the ``parameters" of the NN -- weights and biases -- that are varied during the training to minimize the loss function).

Through trial and error, we found that many of the hyperparameter choices made in  \cite{Kasieczka:2017nvn} could be improved. (A proper scan of hyperparameters would have been ideal but this requires a GPU cluster which we did not have access to.)
While many of these choices are more art than science, and while the best choice may depend heavily on the particular problem domain (e.g.\ the choice that may be ideal for natural images may not be the best choice for jet images), there is some accumulated lore from the field of deep learning about best practices. In this section we will briefly go over some of this lore and explain how its application to jet tagging can significantly improve the DeepTop tagger performance. While we do make an attempt at a somewhat self-contained treatment, we do not promise to have succeeded. We refer the interested  reader to \cite{Nielsen,Goodfellow-et-al-2016} for any missing definitions and more background material.

\subsection{Loss function}\label{Loss function}

In any neural network, the goal is to minimize a ``loss function" $L$ over the NN parameters $\theta$:
\beq
L = \sum_i f(a(\theta,x_i),y_i)
\eeq
The loss function  quantifies how well the network is performing. Here $a(\theta,x_i)$ is the NN prediction and is a function of the NN parameters as well as the input $x_i$ (the jet image in our case); $y_i$ is the truth label of example $i$; and $i$ is summed over all the examples in the training sample. For binary classification problems such as top tagging, we can take $y_i=0$ for signal (tops) and $y_i=1$ for background (not-tops). 

In DeepTop, $f$ was taken to be the mean-squared-error (MSE) $f(a,y)=(a-y)^2$. However, a better choice in classification problems (that we opt for here) is the  cross entropy $f(a,y)=-(y\log a+(1-y)\log(1-a))$. Theoretically speaking, MSE is more appropriate and mathematically/statistically sound for Gaussian random variables, while binary cross entropy is more appropriate for discrete (logistic) classification. 
In more practical terms, using the binary cross entropy for classification tends to avoid the problem of learning slowdown when the predictions are close to zero or one. For more discussion of this see \cite{Nielsen}.

\subsection{Optimizer algorithm}\label{Optimizer}

Having chosen a loss function, we next need to decide on which algorithm we use to minimize it. 
The loss function surface of multilayered NNs is typically non-convex and high-dimensional with multiple flat regions and local minima. So the process of training the NN is highly nontrivial. A poor choice of the optimizer can lead to many undesirable outcomes.

Generally, the optimizers used to train deep networks are based on the idea of gradient descent, where the parameters of the NN are updated according to the derivative of the loss function: 
\beq
\Delta \theta = - \eta \nabla L
\eeq
The learning rate $\eta$ is a hyperparameter that needs to be tuned: gradient descent would take too many steps if $\eta$ is too small, but if $\eta$ is too large one may never converge to a minimum.

Computing the gradient of the full loss function (i.e.\ summed over the entire training set) -- referred to as batch gradient descent -- is generally too time consuming. Instead, most optimizers for deep learning involve some form of Stochastic Gradient Descent (SGD), where the training sample is divided into ``minibatches", and gradients of the loss function are computed on each minibatch. Stepping through the training sample minibatch by minibatch and updating the weights at each step is  then referred to as a ``training epoch".
 While this would appear to provide noisy and inaccurate estimates of the gradient, it actually has many benefits. For instance, introducing some noise into the gradient calculation can prevent the optimizer from becoming stuck in a poor local minimum. Also, while some minibatches may be inaccurate and lead to faulty updates, taken together their cumulative effect can actually greatly speed up the rate of convergence. See \cite{Goodfellow-et-al-2016} for a in-depth discussion of this.

Finally, it is well-known that SGD is very sensitive to the learning rate and other hyperparameters, and optimizing its performance usually requires an in-depth scan and tuning over these quantities (see e.g. \cite{2012arXiv1206.1106S} for a discussion). Therefore, popular alternatives in deep learning are optimizers such as AdaDelta \cite{adadelta} and Adam \cite{adam} that attempt to adaptively determine the optimal learning rate for each parameter and each training epoch. These adaptive versions of SGD usually require little or no manual tuning of a learning rate and are rather insensitive to noisy or large gradients, different model architectures and selection of hyperparameters, etc.

In \cite{Kasieczka:2017nvn}, the optimizer was taken to be vanilla SGD with a minibatch size of 1000 and a fixed learning rate of $\eta=0.003$. These hyperparameters do not appear to have been tuned. Therefore it is not surprising that switching to AdaDelta (with the default settings in Keras \cite{keras}) improves the outcome of training by a considerable amount.\footnote{We also tried using Adam and found very similar improvements.} We obtained further improvements with a slightly reduced learning rate (0.3 instead of 1) and a learning rate schedule (decreasing the learning rate by $1/\sqrt{2}$ when the validation loss does not decrease by more than 0.0005) as compared to the Keras defaults.

We also found a very significant benefit to training with a smaller minibatch size than was used in the DeepTop paper  (128 instead of 1000).\footnote{Perhaps an even smaller minibatch size would help even more, but here we were limited by computation time.} This is in line with the small-to-moderate minibatch sizes ($\lesssim \CO(10^2)$) that are typically  used  in the machine learning literature.
Smaller minibatches give noisier estimates of the gradient, and 
as noted above, this is actually beneficial in non-convex optimization, given that it could push the solution out of the saddle points and shallow minima of the loss function.

\subsection{Architecture}\label{Architecture}
 
Finally, there are  myriad choices involved in specifying the architecture of the neural network. Here we found that the architecture of the DeepTop CNN seemed to be optimal in terms of the number of layers and filter size. But augmenting it with more feature maps (64-128 instead of 8) and  more nodes on dense layers (256 instead of 64) improved the performance considerably.

\begin{figure*}
{%
  \includegraphics[width=1.0\linewidth]{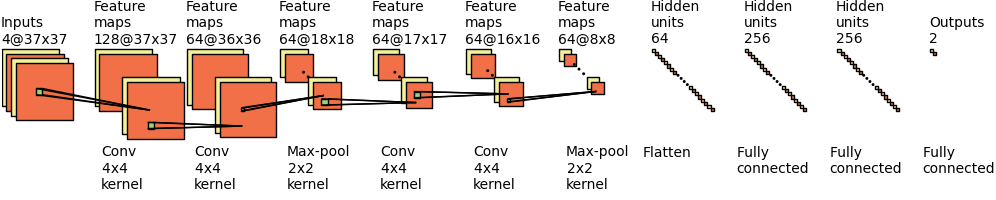}%
}
\caption{\small{Architecture of our CNN top tagger.
}}
\label{fig:NNarch}
\end{figure*}

Our NN architecture is shown in fig.~\ref{fig:NNarch}. The input layer is given by an image of $37\times37$ pixels with (up to) 4 colors.  Next, we define a convolutional layer of 128 feature maps with a $4\times4$ kernel followed by a second convolutional layer of 64 feature maps and similar kernel.\footnote{The larger number of initial feature maps aims to capture all the possible lower lever features of the images. In computer vision applications these features are different shapes (lines, edges, curves, etc.) that the NN uses to build up to higher-level concepts. Although there is not a direct correspondence between typical computer vision images and our images given that jet images are sparse, raising the number of initial feature maps improved the classification accuracy.} Then we have a max-pooling layer with a $2\times2$ reduction factor. Next we apply two more consecutive convolutional layers with 64 features maps with a $4\times4$ kernel each, followed by a max-pooling layer with a $2\times2$ reduction factor. As in \cite{Kasieczka:2017nvn}, we use zero-padding in each convolutional layer to make sure we are not subject to boundary effects. We flatten the 64 maps of the last pooling layer into a single one that is passed into a set of three fully connected dense layers of 64, 256 and 256 neurons each. (Restricting the first dense layer to 64 neurons was motivated by practical considerations. It keeps the number of weights at a manageable level, speeding up training time and ameliorating overfitting.) Finally, the last dense layer is connected to the output layer of 2 neurons which produces the probability that the jet originated from a top or not. We use rectified linear units (ReLU) as the activation functions on all the layers, except for the output layer where we use the softmax function. Also, our final training sample was large enough so that regularization techniques, such as dropout, were not necessary. 

The neural network is implemented on an NVidia Tesla K80 GPU using the NVidia CUDA platform (CUDA drivers, toolkit and cuDNN library).
The code for the CNN is written in Python, using the deep learning library Keras \cite{keras} with the TensorFlow \cite{tensorflow2015-whitepaper} backend. The weights are initialized with the Keras default settings. 

We arrived at the NN architecture used in this paper mainly by trial and error. Due to limited resources, a thorough scan of NN architectures was not possible, however this would obviously be desirable. It is easily possible that further performance gains could be obtained with such a scan.\footnote{We note that a limited scan was carried out in the DeepTop paper. However, they only considered 6, 8 and 10 feature maps per convolutional layer, which does not include the 64-128 feature maps used in this work.}

\section{Image preprocessing}\label{Preprocessing}

In the original DeepTop paper \cite{Kasieczka:2017nvn}, the image preprocessing steps were found to actually {\it decrease} the performance of the tagger. This is surprising since usually preprocessing improves classifier performance. 

The DeepTop preprocessing steps were as follows. First they pixelated the image according to their detector resolution. Then they shifted such that the maximum pixel intensity as defined by a 3x3 window was at the origin. Next, they rotated such that the second maximum was in the 12 o'clock position, and they flipped to ensure that the third maximum is in the right half plane. 
Finally, they normalized each image so that the pixel intensities are between 0 and 1.

\begin{figure}[t]
\begin{center}
\includegraphics[width=0.6\linewidth]{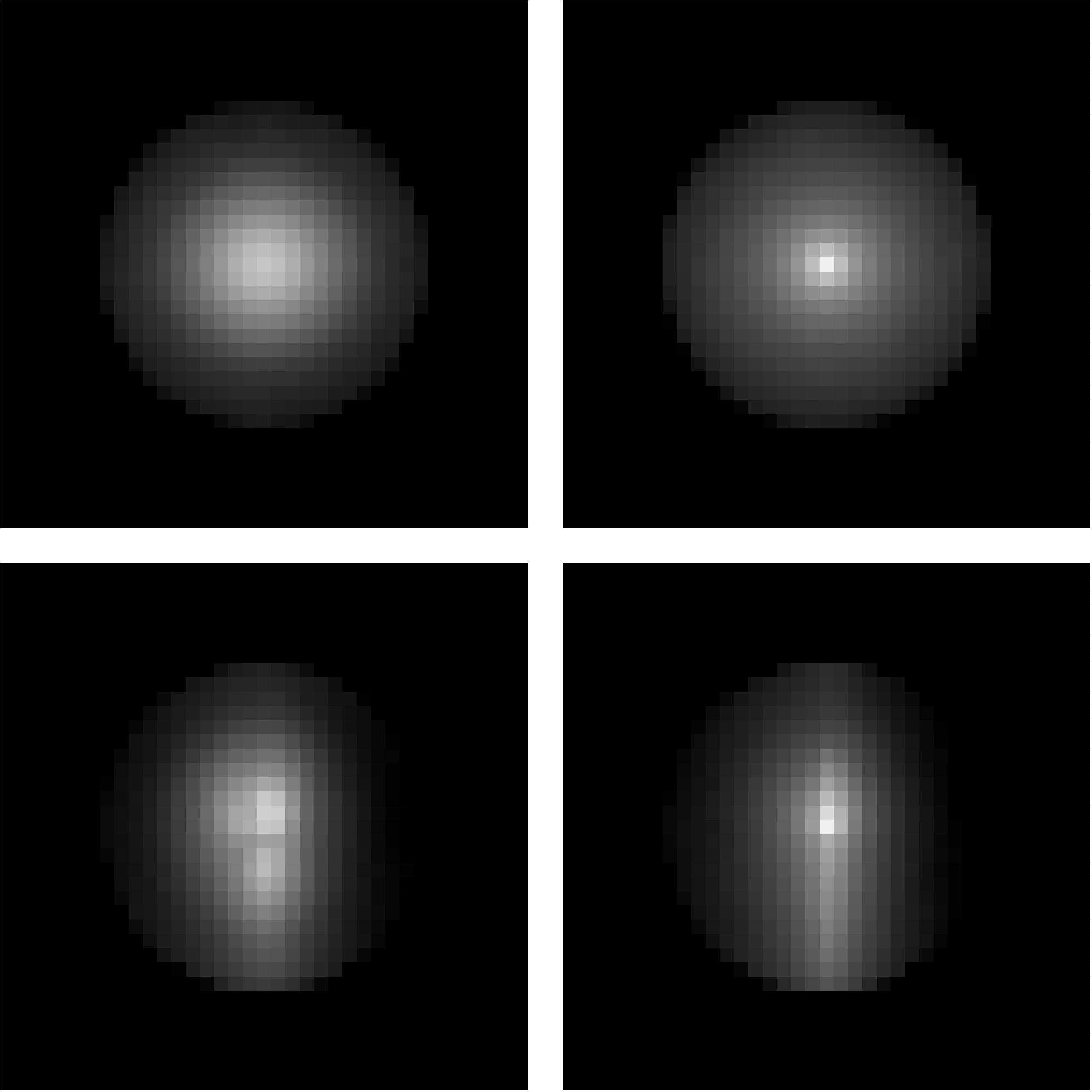}
\caption{The average of 100k jet images drawn from the CMS sample ($37\times 37$ pixels spanning $\Delta\eta=\Delta\phi=3.2$).  The grayscale intensity corresponds to the total $p_T$ in each pixel. Upper: no preprocessing besides centering. Lower: with full preprocessing. Left: top jets. Right: QCD jets}
\label{fig:avgimg}
\end{center}
\end{figure}

Our preprocessing steps differ from this in the following ways. First of all, we perform all preprocessing {\it before} pixelating the image. This makes the most sense for the CMS sample which separates the much-higher-resolution tracks from the calorimeter towers. But it also appears to have some benefit even for the calo-only jets of the DeepTop sample. Our first step is to calculate the $p_T$-weighted centroid of the jet and the $p_T$-weighted principal axis. Then we shift so that the centroid is at the origin and we rotate so that the major principal axis is vertical. In contrast to DeepTop, we flip along {\it both} the L-R and the U-D axes so that the maximum intensity is in the upper right quadrant. Finally, after doing all these transformations, we pixelate the image and then normalize it to unit total intensity (i.e.\ divide by the total $p_T$). 

To demonstrate the effectiveness of our preprocessing steps, we show in fig.~\ref{fig:avgimg} the average of 100k top and QCD jet images drawn from the high $p_T$ CMS jet sample, with and without preprocessing. Although below we consider color images where the track $p_T$'s and neutral $p_T$'s are considered separately, here we restrict ourselves to grayscale images where they are added together. We see that even without preprocessing, the average images are quite different, with the QCD jets being much more peaked than the top jets. After our preprocessing steps, the 3-prong substructure of the top jets becomes readily apparent, while the QCD jets remain more dipole-like. (This should be contrasted with the average images in the DeepTop paper, where the 3-prong substructure of the top jets is much less apparent.)

\section{Other improvements}\label{Other improvements}

\subsection{Sample size} \label{Sample size}

\begin{figure}[t]
\begin{center}
\includegraphics[width=0.8 \linewidth]{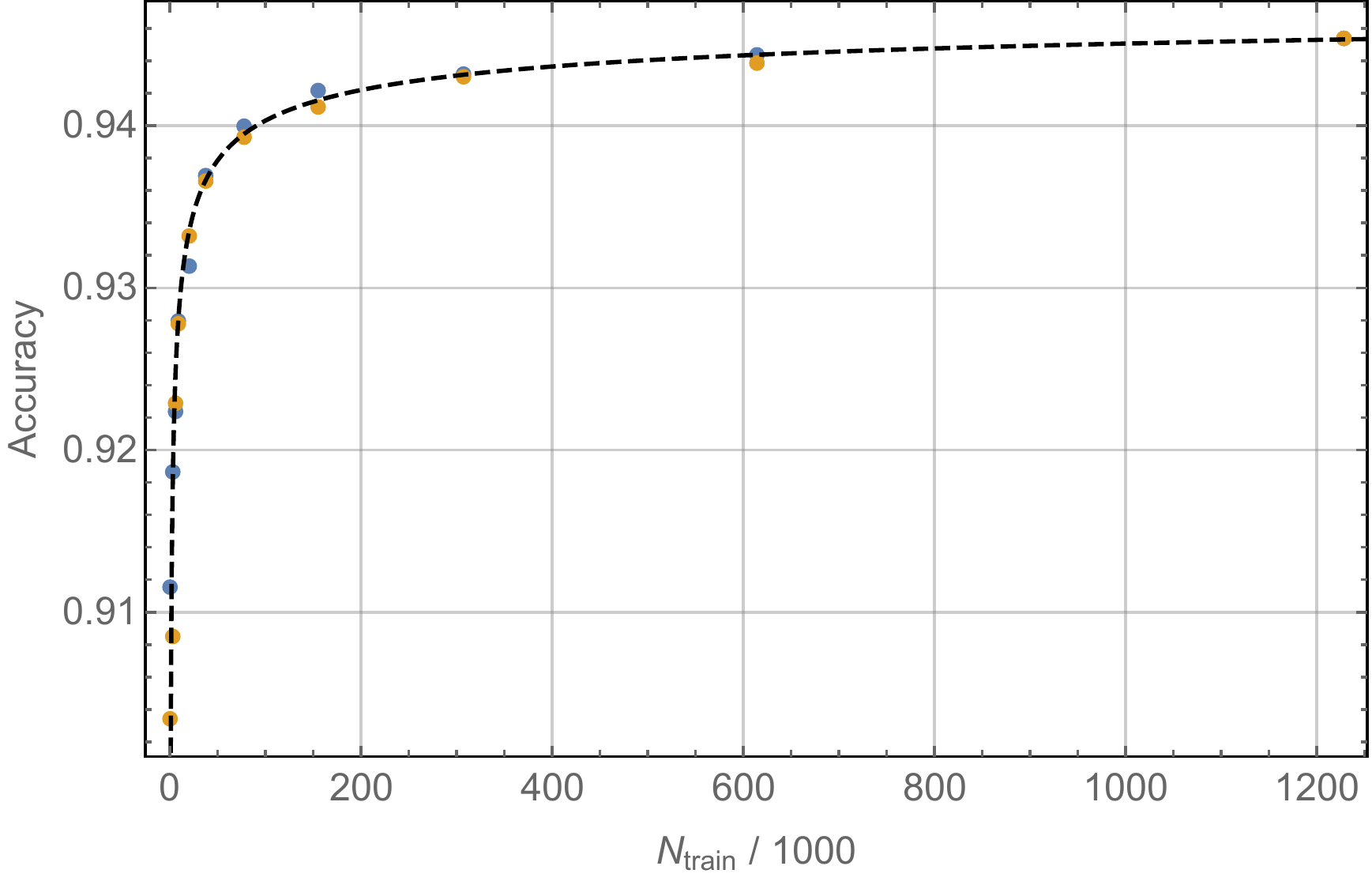}
\caption{In blue (yellow) are the learning curves for the test accuracy vs.\ training sample size for the CMS jets (DeepTop jets). The CNN used is our final tagger but with grayscale images. The learning curve for DeepTop jets has been shifted up by a constant offset of 0.075. Shown also in black, dashed is a heuristic least-squares fit to an inverse power law with uncertainties given by $1/\sqrt{N_{train}}$.}
\label{fig:samplesizescan}
\end{center}
\end{figure}

In the DeepTop paper, the training samples were limited to 150k+150k. Here we explore the effect  on our CNN top tagger of increasing the training sample size. Shown in fig.~\ref{fig:samplesizescan} are the learning curves for the test accuracy vs.\ training sample size, for our two different jet samples. (The training sample size is defined to be the number of top jets in the training sample; an equal number of QCD jets were used. The test sample size was fixed at 400k+400k jets.) We have shifted the learning curve for the DeepTop sample by a constant 0.075; interestingly, it lines up almost perfectly with the learning  curve for the CMS sample. This is evidence that the shape of the learning curve is independent of the fiducial jet selection (although the asymptotic value clearly depends strongly on it). In any event, we see that the performance is basically saturated for $\gtrsim 1$M jets (for our final CNN tagger, we train on 1.2M+1.2M jets). 

We also indicate in  fig.~\ref{fig:samplesizescan} the result of a least-squares fit  of an inverse power law  $a+b/N_{train}^c$ to the learning curve. This description of the learning curve may be a general empirical feature of machine learning \cite{learning-curve}. However, lacking a precise understanding of the uncertainties on the test accuracies (the sample variance from both the test set and the training set contribute), we cannot provide a detailed description of the fit. 
Here, to perform the fit, we estimated the uncertainty on each value of the test accuracy using a simple $1/\sqrt{N_{train}}$ scaling.\footnote{We have tested this scaling using a small number of pseudoexperiments for small values of $N_{train}$ and it appears to hold.} We merely include this fitting function to guide the eye. One sees visually that it seems to describe the learning curves well.

\subsection{Color}\label{Color}
 
Inspired by \cite{Komiske:2016rsd}, we also added color to our images from the CMS sample. (The DeepTop sample was calo-only so we could not add color to them.) The four colors we used were neutral and track $p_T$ per pixel, the raw number of tracks per pixel, and the number of muons per pixel. The last color was not considered in \cite{Komiske:2016rsd}, which focused on quark vs.\ gluon tagging. Obviously, muons can be considered a crude proxy for $b$-tagging and should play a role in any top tagger.  (For more comments on $b$-tagging, see Section \ref{Outlook}.) 

Interestingly, we found that adding color to the images led to significant overfitting for smaller training sample sizes. Evidently, while the color adds information to the images, it also increases the noise, and with too few training examples, the network learns to fit the noise. This problem went away when the training sample was increased to 1.2M+1.2M, which is why we choose to place the color improvement last. 
 
\begin{table*}[t]
\begin{tabular}{|c|c|c|}
\hline
& DeepTop minimal & Our final tagger  \\ \hline
\multirow{4}{*}{Training} & SGD & AdaDelta \\& $\eta=0.003$ & $\eta=0.3$ with annealing schedule\\& minibatch size=1000& minibatch size=128\\& MSE loss & cross entropy loss
\\\hline
\multirow{2}{*}{CNN architecture}
& 8C4-8C4-MP2-8C4-8C4- & 128C4-64C4-MP2-64C4-64C4-MP2- \\ &64N-64N-64N &  64N-256N-256N
\\\hline
\multirow{2}{*}{Preprocessing} & pixelate$\to$center & center$\to$rotate$\to$flip\\ & $\to$ normalize & $\to$ normalize$\to$pixelate
\\\hline
Sample size & 150k+150k & 1.2M+1.2M
\\\hline
Color & $p_T^{calo}=p_T^{neutral}+p_T^{track}$ &  $(p_T^{neutral},  \, p_T^{track},\, N_{track},\, N_{muon})$
\\\hline
\end{tabular}
\caption{Summary of our final CNN tagger, together with the original DeepTop tagger.
\label{tab:finaltagger}}
\end{table*}

\section{Final comparison}\label{Final comparison}

The full specifications of our final tagger are summarized in table \ref{tab:finaltagger} side-by-side with those of the original DeepTop tagger.

\begin{figure}[t]
\begin{center}
\includegraphics[scale=0.8]{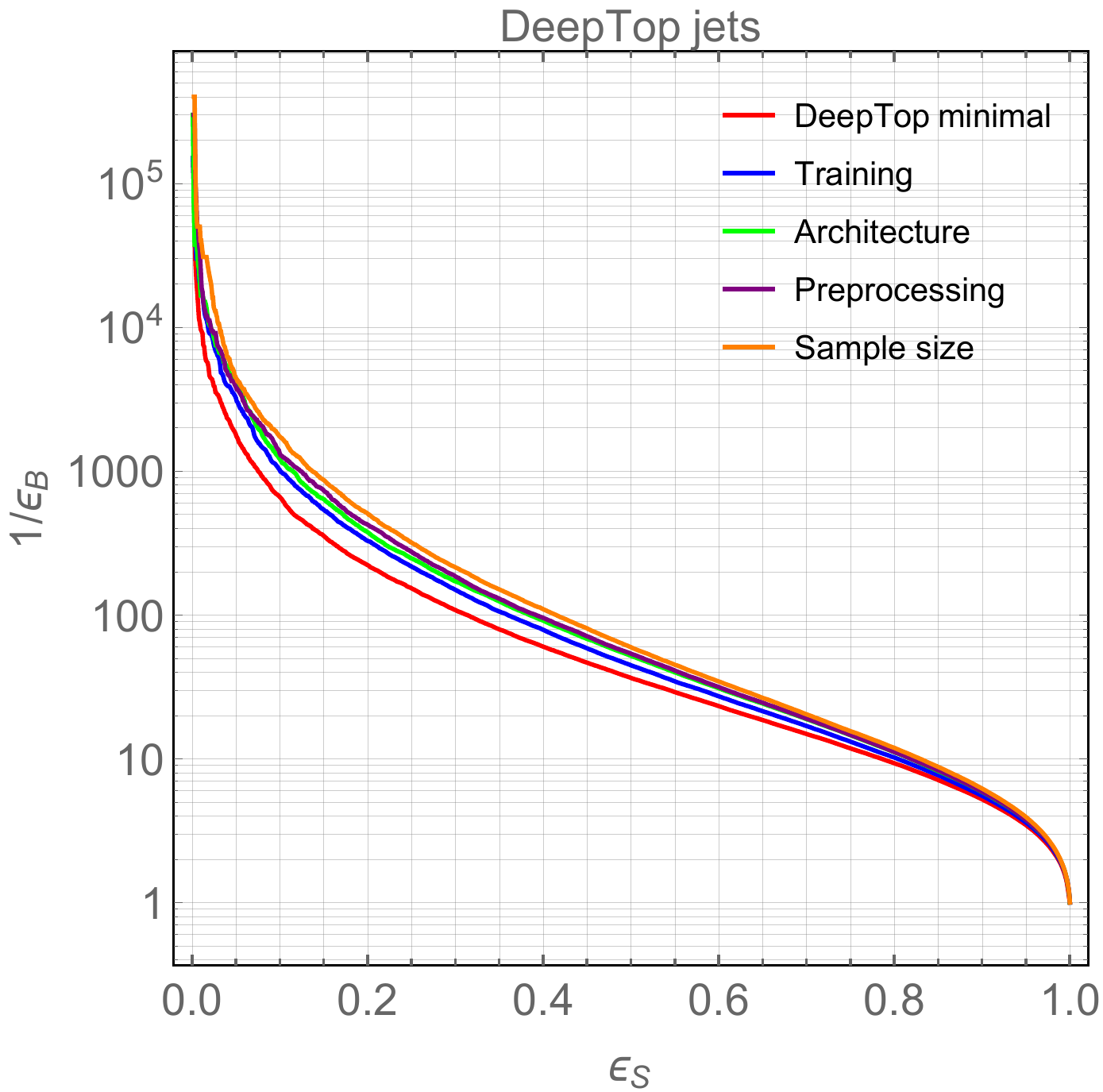}
\caption{Sequence of ROC curves (background rejection $1/\epsilon_B$ vs.\ tagging efficiency $\epsilon_S$) illustrating the cumulative effects of the various improvements to the DeepTop tagger, for the DeepTop jet sample. Our final tagger including all the improvements is shown in orange.
}
\label{fig:deeptopimprovements_deeptopjets}
\end{center}
\end{figure}

\begin{figure}[t]
\begin{center}
\includegraphics[scale=0.8]{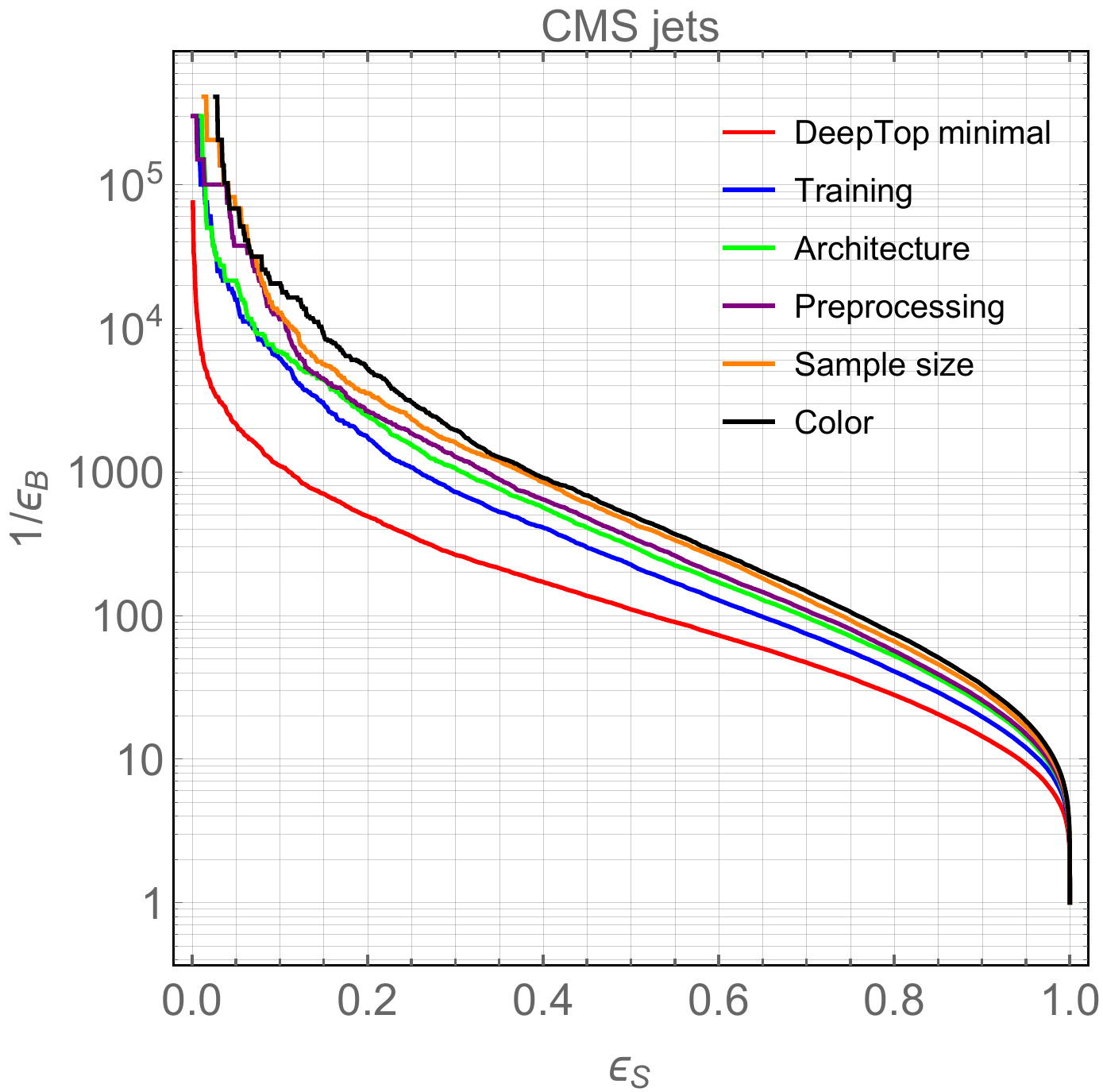}
\caption{Same as fig.~\ref{fig:deeptopimprovements_deeptopjets} but for the CMS jet sample.
}
\label{fig:deeptopimprovements}
\end{center}
\end{figure}

Having gone through all the improvements (loss function, optimizer, CNN architecture, image preprocessing, sample size and color) to the DeepTop tagger in the preceding sections, we are now ready to put them all together and quantify their cumulative effects on the tagger performance. 
Shown in  figs.~\ref{fig:deeptopimprovements_deeptopjets}--\ref{fig:deeptopimprovementsratio} and table~\ref{tab:metrics} are ROC curves and aggregate metrics characterizing these effects.
The baseline in these plots is always the DeepTop minimal column in table \ref{tab:finaltagger}, applied to the two different jet samples in table \ref{tab:jetsamples}. Each modification is then added cumulatively to this baseline. Here is a more detailed breakdown (each entry here corresponds to moving from left to right sequentially in the corresponding category of table \ref{tab:finaltagger}):

\begin{itemize}

\item The end result of all of our improvements to the training (loss function and optimizer) is the blue curves in figs.~\ref{fig:deeptopimprovements_deeptopjets}-\ref{fig:deeptopimprovementsratio}. This gave the single largest boost to the performance of all the different modifications we considered. Furthermore, we find that over half of the improvement here is due solely to the smaller minibatch size. We also note in passing that the better training methods allowed us to vastly  speed up the training time, as we only need $\CO(10)$ training epochs to converge instead of the $\CO(10^3)$ epochs of the DeepTop paper. 

\item Improving the DeepTop architecture with more feature maps and more nodes on hidden layers brought about another substantial gain in performance, this is indicated in the green curves in figs.~\ref{fig:deeptopimprovements_deeptopjets}-\ref{fig:deeptopimprovementsratio}.

\begin{figure}[t!]
\begin{center}
\includegraphics[scale=0.5]{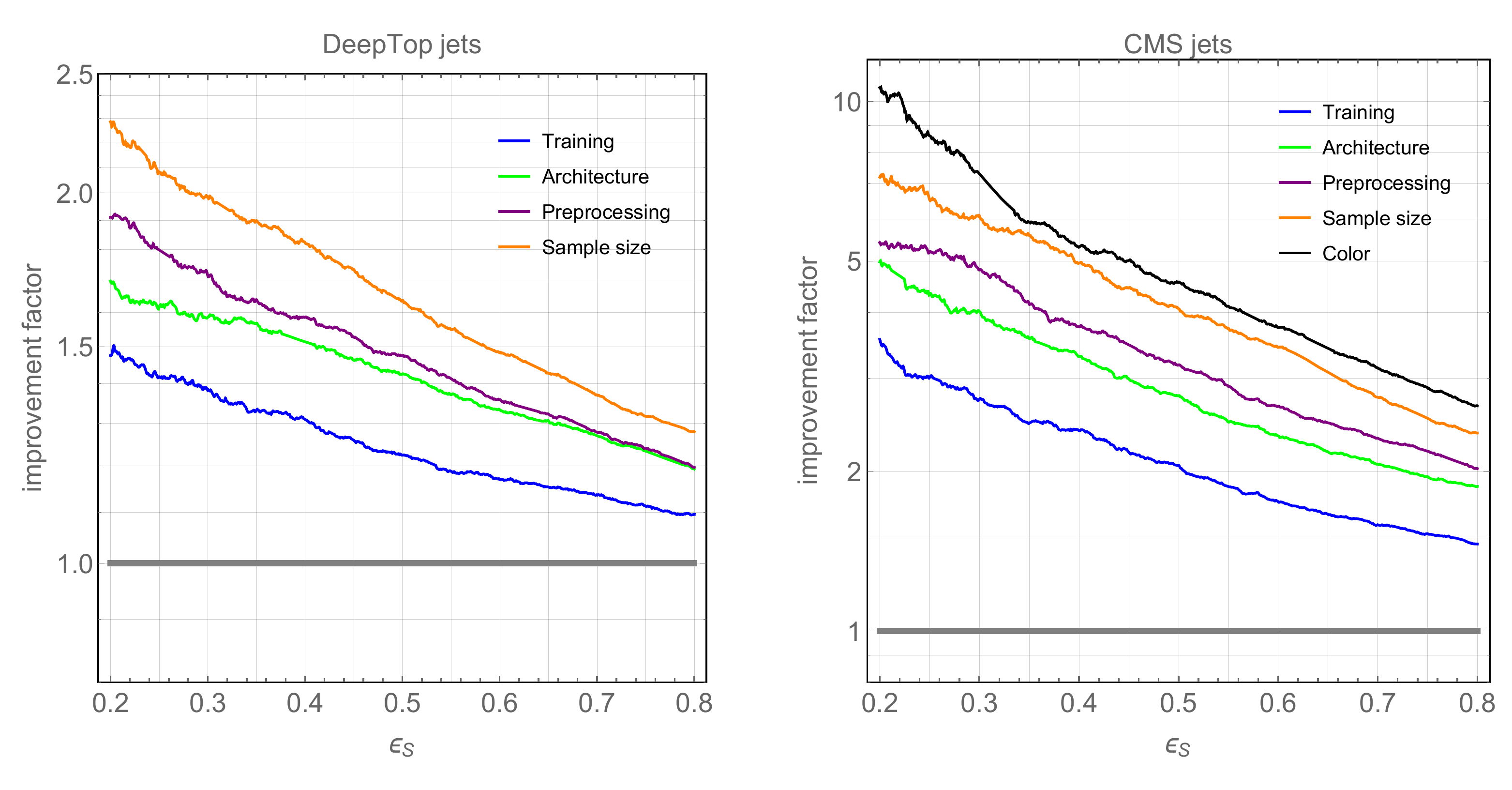}
\caption{Ratio of the ROC curves in figs.~\ref{fig:deeptopimprovements_deeptopjets}--\ref{fig:deeptopimprovements} over the minimal DeepTop tagger ROC curve, providing another view of the cumulative improvements. 
}
\label{fig:deeptopimprovementsratio}
\end{center}
\end{figure}

\begin{table*}[h!]
\begin{tabular}{|c||c|c||c|c|}
\hline
& \multicolumn{2}{|c|}{DeepTop jets} & \multicolumn{2}{|c||}{CMS jets}  \\
\hline\hline
Improvement & Accuracy & AUC & Accuracy & AUC \\ \hline\hline
Baseline  & 85.5\% & 0.930 & 91.7\% & 0.975
\\\hline
Training  & 86.1\% & 0.935 & 93.4\% & 0.983
\\ \hline
Architecture  & 86.6\% & 0.939 & 94.0\% & 0.985
\\\hline
Preprocessing  & 86.7\% & 0.940 & 94.2\% &  0.986
\\\hline
Sample Size  & 87.0\% & 0.943 & 94.5\% & 0.988
\\\hline
Color   & --- & --- & 94.8\% & 0.989
\\
\hline
\end{tabular}
\caption{Accuracy and area under the curve (AUC) of our tagger after adding the modifications over DeepTop minimal.
\label{tab:metrics}}
\end{table*}

\item The result of our image preprocessing steps is a (relatively modest) improvement in tagger performance, as indicated by the purple curves in figs.~\ref{fig:deeptopimprovements_deeptopjets}-\ref{fig:deeptopimprovementsratio}.

\item We found that increasing the training sample size by a factor of $\sim 10$ significantly improved the performance. The improvement using 1.2M+1.2M jets (which according to fig.~\ref{fig:samplesizescan} is enough to saturate the best-possible performance of this tagger) is indicated by the orange curves in figs.~\ref{fig:deeptopimprovements_deeptopjets}-\ref{fig:deeptopimprovementsratio} (the previous ROC curves were based on the DeepTop training sample size of 150k+150k jets).

\item Adding color (only possible for the CMS jet sample that differentiates tracks from neutrals) resulted in a very modest improvement in the tagger performance, shown in the black curve in figs.~\ref{fig:deeptopimprovements}-\ref{fig:deeptopimprovementsratio}. 

\end{itemize}

We see that with these modifications we can achieve a factor of $\sim 3$--10 improvement (depending on the tagging efficiency) in the background rejection rate for the CMS jet sample and a factor of $\sim 1.5$--2.5 improvement for the DeepTop jet sample. 

It is interesting that the improvements are much greater for the CMS jet sample than the DeepTop jet sample. Perhaps the tops vs.\ QCD jets in the CMS  sample have more subtle distinguishing features that can only be learned with the improved methods. Regardless of the reason, this comparison illustrates the strong effect that the fiducial jet selection can have on tagger performance.
And although our improvements are more modest for the DeepTop sample, they still do improve it by a factor of $\sim 2$, which is still quite significant. This demonstrates that the  principles described in the previous subsections which motivated these improvements do have general validity.

\begin{figure}[t]
\begin{center}
\includegraphics[width=0.7\linewidth]{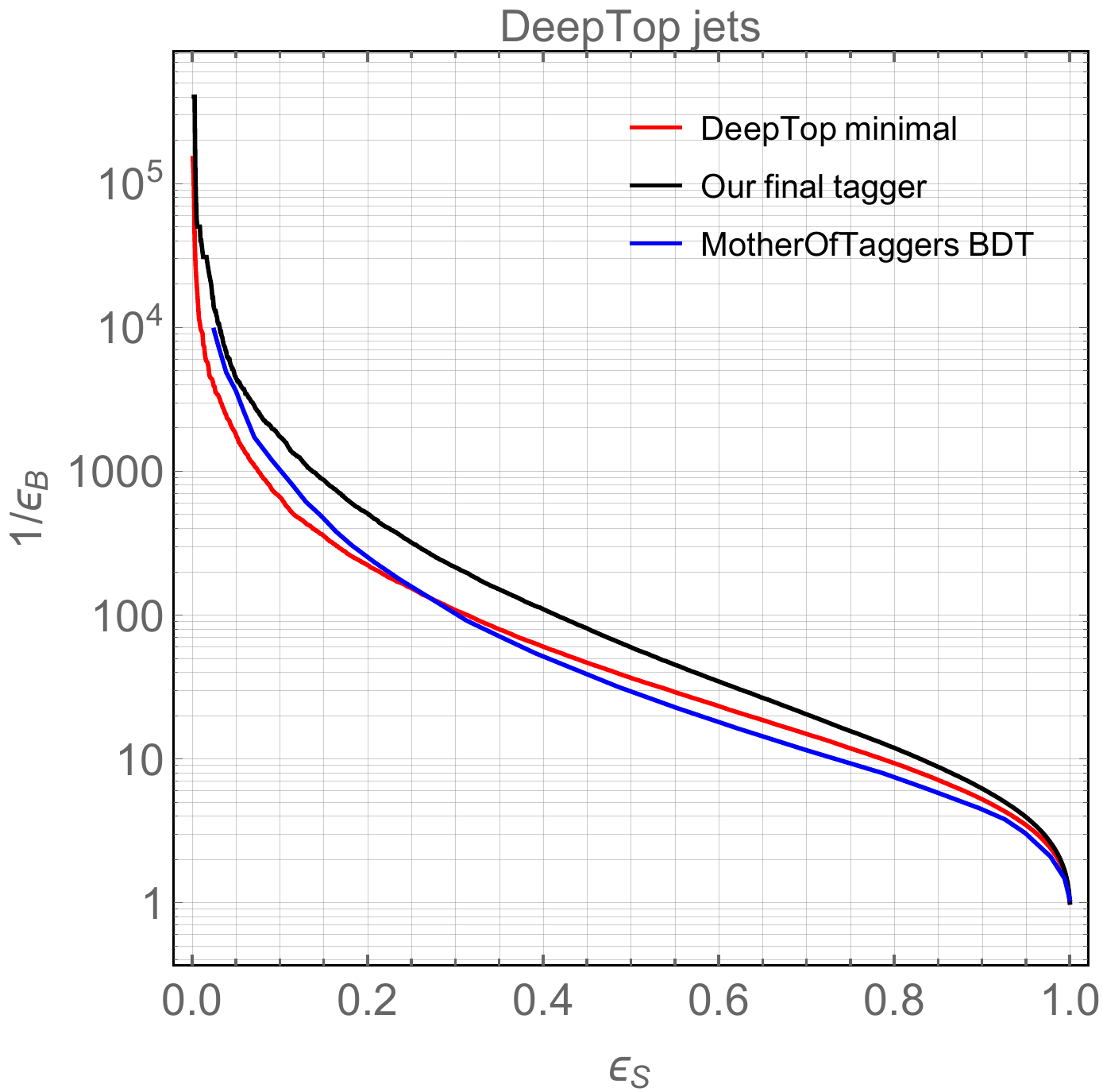}
\caption{ROC curves comparing our best top tagger (black), the original DeepTop tagger (red), and the ``MotherOfTaggers" BDT built out of high-level inputs from \cite{Kasieczka:2017nvn} (blue solid), for the DeepTop jet sample.}
\label{fig:finalcomparison_deeptopjets}
\end{center}
\end{figure}

\begin{figure}[t]
\begin{center}
\includegraphics[width=0.7\linewidth]{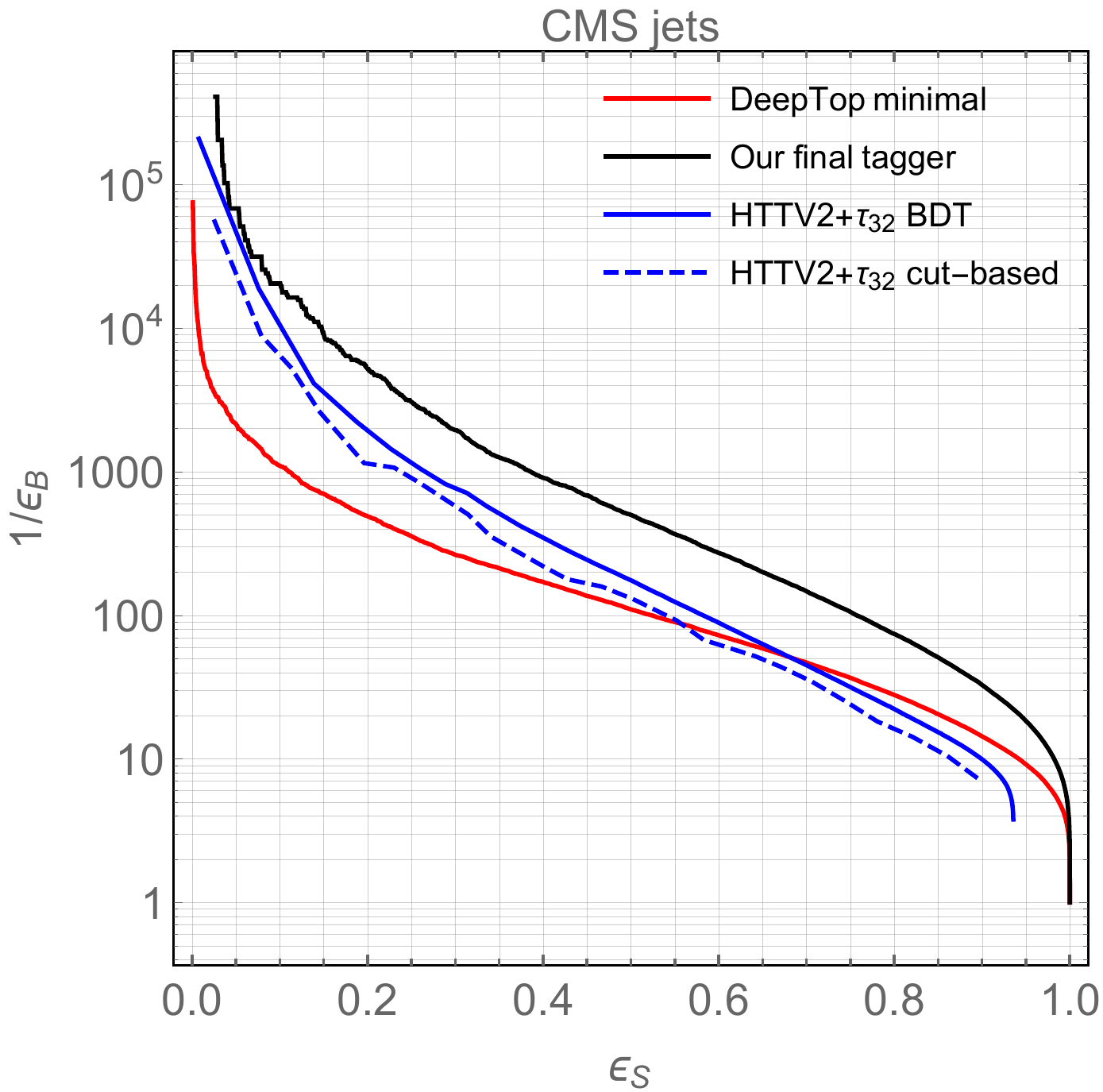}
\caption{ROC curves comparing our best top tagger (black), the original DeepTop tagger (red), the cut-based top-tagger from \cite{CMS:2016tvk} using variables from HTTV2 and $\tau_{32}$ (blue dashed), and a BDT built out of those same variables (blue solid), for the CMS jet sample.}
\label{fig:finalcomparison}
\end{center}
\end{figure}

The comparison between our tagger and state-of-the-art conventional top taggers that use high-level features is shown in fig.~\ref{fig:finalcomparison_deeptopjets} for the DeepTop jet sample and in fig.~\ref{fig:finalcomparison} for the CMS jet sample. As discussed in Section \ref{Methodology}, for the DeepTop jet sample, we compare directly against their ``MotherOfTaggers" BDT ROC curve (i.e.\ without recasting it). For the CMS jet sample, we include two taggers that are representative of the  state-of-the-art in top-tagging with high-level features: a cut-based top-tagger using variables from HTTV2  and N-subjettiness, and a BDT built out of those same variables. The BDT is trained on the same 1.2M+1.2M jets as our final CNN tagger. The BDT  improves the performance of the high-level cut-based tagger by a moderate amount. 

For the DeepTop jet sample, the baseline tagger was already comparable to the BDT, and our improvements to the former raise it above the BDT by a factor of $\sim 2$. Meanwhile, for the CMS jet sample, it is surprising to see  that the baseline tagger is outperformed by even a simple  cut-based tagger at lower tag efficiencies. This again highlights the importance of optimizing a tagger for each fiducial jet selection. Thanks to the factor of 3--10 improvement over the baseline, our final CNN top tagger still shows substantial gains (a factor of $\sim 3$ in background rejection) compared to the BDT. One exception to this is at the lowest tag efficiencies ($\epsilon_S\sim 0.1$), where the BDT and the deep learning tagger perform very similarly (this can be seen also in the DeepTop sample). This could be because at these low tag efficiencies, the top is very easy to identify and discriminate against QCD, and so the gain from deep learning is minimized.

\section{Outlook} \label{Outlook}

In this paper, we showed for the first time how a top tagger based on deep learning and low-level inputs (raw jet images) can significantly outperform state-of-the-art conventional top taggers based on high level inputs. Taking the DeepTop tagger as a starting point, we explored a number of modifications, most of them quite simple, that overall improve the performance  by up to a factor of $\sim 10$ in the ROC curve. Compared to a BDT trained on high-level inputs, our image-based deep-learning top tagger performs better by as much as a factor of $\sim 3$.

We believe our work illustrates the enormous promise and potential of modern machine learning. Many more exciting results are sure to follow. In this section we will briefly discuss some of the interesting future directions.

In this work, we made various simplifying assumptions that should be relaxed in future studies. For instance, we ignored pileup. This was motivated by the fact that these are very high $p_T$ jets and we are just trying to classify, instead of trying to measure anything precisely, so we expect pileup to have a negligible effect. But this should be checked -- for any actual experimental application one would want to demonstrate the performance of the tagger under realistic pileup conditions.  We also restricted to two narrow ranges (350-450~GeV and 800-900~GeV) of top $p_T$s. The stability of a tagger performance under a broad range of $p_T$s is important to experimentalists, to avoid artificially sculpting the data. 

Another glaring omission is $b$-tagging. Here we have just relied on the momentum four-vectors of the jet constituents, and have not used any impact parameters, displacements or secondary vertex finding.  Obviously, since this information is orthogonal to the momenta, we expect that adding $b$-tagging will give a further boost to the tagger performance. It would be interesting to know whether this boost is enhanced by deep learning or not. 

The reason we were not able to add $b$-tagging is because there is not enough publicly available information to accurately recast the secondary vertex finders used by the experimental collaborations, or even the impact parameters (IPs). The IP resolutions have not been updated past 7 TeV \cite{CMS-PAS-BTV-11-002}, and they are for single isolated tracks or at best very low $p_T$ tops. These IP resolutions are likely to be unrealistically good for tracks in high $p_T$ boosted top environments. Indeed,
when we attempted to implement IP significance $b$-tagging (say, along the lines of \cite{Rizzi:927385}) using these publicly available IP resolutions,  we found too large of an improvement to the top tagger performance compared to what one sees e.g.~in \cite{CMS:2016tvk}.

Another relevant topic that we have not explored in this paper concerns the issue of overtraining on MC simulations. Clearly, our tagger has no problem generalizing from the training sample to the test sample, but the question is how representative this sample is of the actual data. Since we only used \textsc{Pythia} \cite{Sjostrand:2014zea}  with some default settings, this question remains unanswered. Some have tried to address it using \textsc{Herwig} \cite{Bahr:2008pv,Bellm:2015jjp} as a stand-in for the data (i.e.\ training on \textsc{Pythia} jets and then testing on \textsc{Herwig} jets to see if there is any degradation in performance), but this is most meaningful if somehow \textsc{Herwig} is more realistic than \textsc{Pythia}. Otherwise any conclusions from \textsc{Pythia} vs.~\textsc{Herwig} comparisons could be misleading.

As noted above, we did not have access to a GPU cluster here. With such computing resources, it would be possible, and important to do a proper architecture and hyperparameter scan to see if the NN performance could be further improved.
Our architecture considered here was inspired by the DeepTop paper. However, there are many state-of-the-art CNN architectures out there such as AlexNet \cite{Krizhevsky}, Fast-R-CNN \cite{DBLP:journals/corr/Girshick15}, VGG \cite{DBLP:journals/corr/SimonyanZ14a}, ResNet \cite{DBLP:journals/corr/HeZRS15}, GoogLeNet \cite{DBLP:journals/corr/SzegedyLJSRAEVR14}, etc. It would be interesting to test these out and see if any of them offer any further benefit.

It should be straightforward  to generalize the top tagger in this work to classify other boosted objects such as $W/Z$ bosons, Higgses, and BSM particles. It would also be interesting to broaden the scope to include partially-merged and fully resolved tops in the list of taggable particles. In this sense, the tagger could have a performance dependent on these two categories, resulting in a greater background rejection at a fixed tag efficiency for merged tops.

Beyond boosted jet tagging, there are countless other potential applications of deep learning to the LHC.  For instance, classification of full events 
is explored in \cite{Louppe:2017ipp}. Furthermore, there are papers that apply Generative Adversarial Networks \cite{2014arXiv1406.2661G} for simulations in high energy physics in \cite{Paganini:2017dwg,deOliveira:2017rwa, Paganini:2017hrr,deOliveira:2017pjk}, where the main purpose is to drastically reduce the event generation time taken by the Geant4 package \cite{Agostinelli:2002hh} to emulate the detector response. Other studies focus on extending the ML based classifiers from fully supervised (each event is labeled as signal or background for training purposes) to weakly supervised 
\cite{Cohen:2017exh,Dery:2017fap, Metodiev:2017vrx, Komiske:2018oaa}. Another interesting direction to explore would be using unsupervised learning to find all the categories (or discover new ones) of boosted objects or other types of signatures. Given all of these interesting future directions (and more), we believe we are just starting to grasp the scope of the many applications of ML in high energy physics.

\section*{Acknowledgements:}
We thank Kostas Bekris, Andrew Dobson, Jingjing Liu, Chaitanya Mitash and Kun Wang for helpful discussions about computer vision and deep learning. We also thank Matt Buckley, Alejandro Gomez Espinosa, Eva Halkiadakis, Angelo Monteux, Kevin Nash, Marc Osherson, Satish Ramakrishna, Scott Thomas, Matthew Walker and especially Gregor Kasieczka for useful discussions about machine learning applications and physics at the LHC. We are grateful to Marco Farina, Gregor Kasieczka, Yuichiro Nakai and Scott Thomas for helpful comments on the manuscript.  Finally, we are especially grateful to the Center for Computational Biomedicine, Imaging and Modeling (CBIM) at Rutgers University and the IAS SNS Linux Compute Servers for providing the computational resources (GPUs) employed in this work. This work was supported by DOE grant DE-SC0010008.


\appendix

\section{Validating our DeepTop implementation}\label{Validation DeepTop}

Here we will validate our implementation of the DeepTop tagger \cite{Kasieczka:2017nvn} that forms the basis of this work. 
Following their specifications, as described in table \ref{tab:jetsamples} (14~TeV collisions,   $350~{\rm GeV}<p_T<450~{\rm GeV}$, $|\eta|<1$, anti-$k_T$ $R=1.5$ calo jets, $\Delta R(t,j)<1.2$ match requirement, no merge requirement, $\Delta\eta\times\Delta\phi=0.1\times 5^\circ$ toy calorimeter, 40$\times$40 pixel images), with the ``minimal" preprocessing option described in their paper (centering only), we produced 600k+600k top and QCD jet images, split 25\%/25\%/50\% into training, validation and test samples as in  \cite{Kasieczka:2017nvn}.

\begin{figure}[t]
\begin{center}
\includegraphics[scale=0.5]{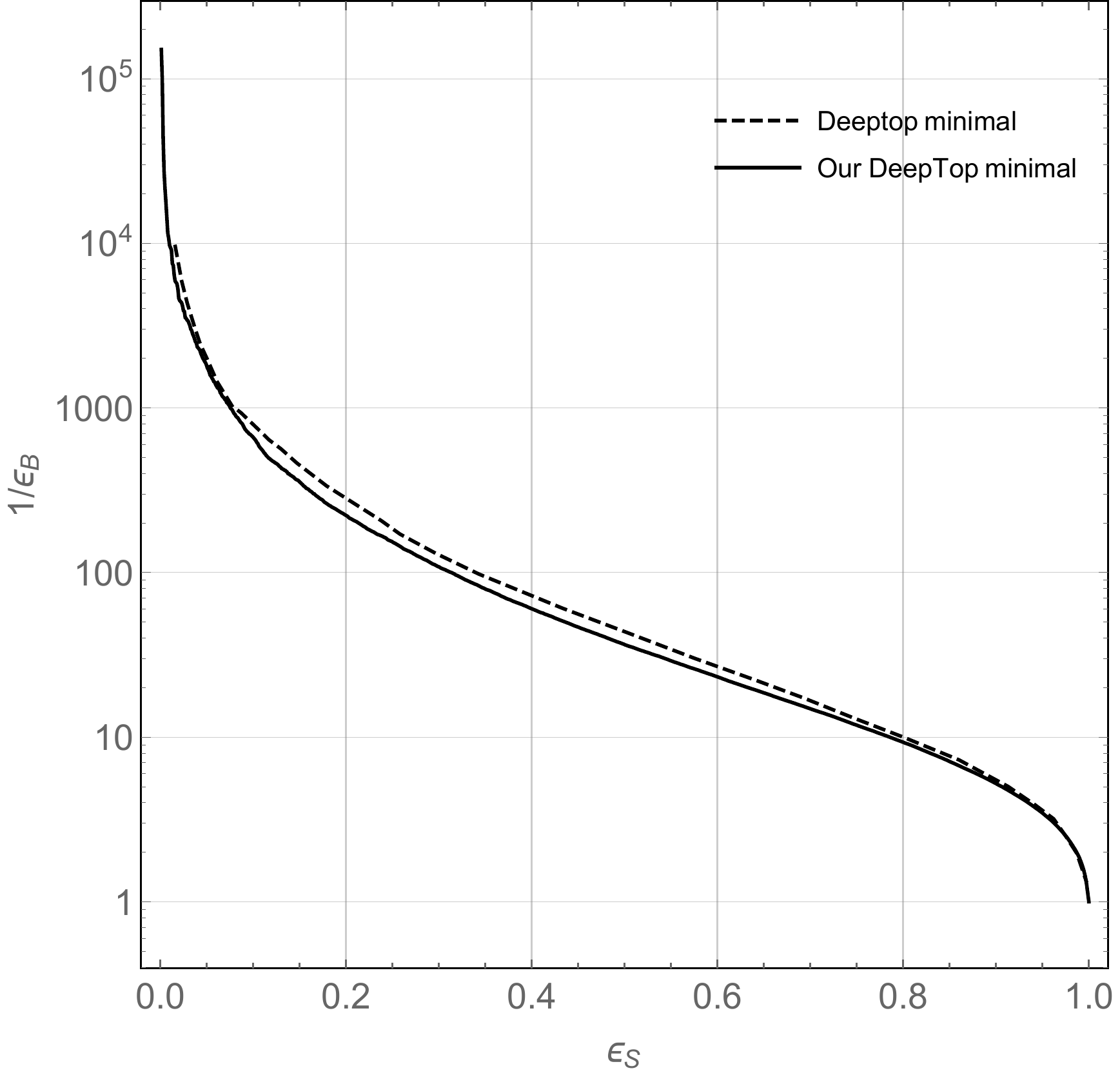}
\caption{ROC curves validating our implementation of the DeepTop tagger with minimal preprocessing (solid) against the original (dashed). The latter was obtained by digitizing the  ``DeepTop minimal" curve in fig.~8 of \cite{Kasieczka:2017nvn}. We see that the agreement is excellent.}
\label{fig:deeptop}
\end{center}
\end{figure}

We used the ``default architecture" shown in fig.~4 of  \cite{Kasieczka:2017nvn}. This, together with the training methods used in the DeepTop paper were described in section \ref{Methodology}. Following these same methods, the result of our validation is shown in fig.~\ref{fig:deeptop}. We see that the agreement is excellent.

\section{Validating our HEPTopTaggerV2 implementation}\label{Validation HEPTopTaggerV2}

Next we turn to validating our implementation of HEPTopTaggerV2 (HTTV2) and Nsubjettiness as used in \cite{CMS:2016tvk}. 
As described in section \ref{Methodology}, their jet samples are in line with our CMS sample, except for some slight differences, specifically $800<p_T<1000$ and $R=0.8$. 

The HTTV2 algorithm  takes the constituents of a jet as input, attempts to cluster them into subjets consistent with a $b$ and a $W$, and outputs a short list of kinematic variables, $m_{jet}$, $f_{rec}$ and $R_{opt}$. The first is the jet mass and obviously should be close to the top mass. The second is a measure of how $W$-like a subjet is. The third is a measure of the optimal jet radius which may be different than the input jet radius.\footnote{For some jets, the HTTV2 may fail to find three or more subjets, 
in which case it produces no outputs. This failure mode must be included in the efficiency calculation of any HTTV2-based tagger. } Finally, the N-subjettiness variables $\tau_i$ are observables built out of the jet constituents that measures how likely the jet is to have a given number of subjets.

\begin{figure}[t]
\begin{center}
\includegraphics[scale=0.5]{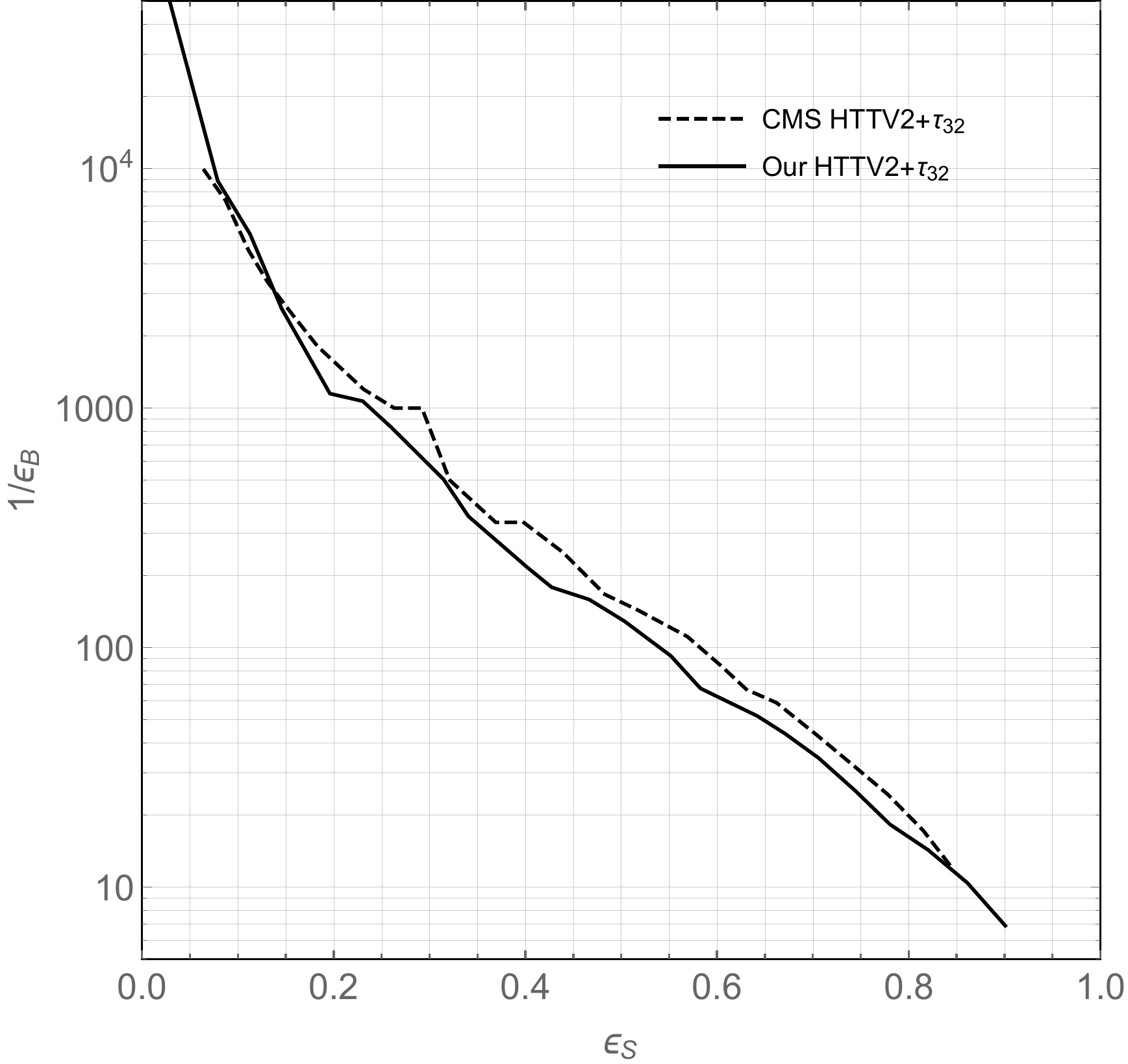}
\caption{ROC curves validating our implementation (solid) of the HTTV2+N-subjettiness cut-based tagger described in \cite{CMS:2016tvk}. The CMS curve (dashed) was digitized from fig.~7R of \cite{CMS:2016tvk}. We see that the agreement is pretty good.}
\label{fig:httv2}
\end{center}
\end{figure}

Using $m_{jet}$, $f_{rec}$ and $\tau_{32}\equiv\tau_3/\tau_2$, CMS scans over simple window cuts to produce the optimal mistag rate for a given tag efficiency. The resulting ROC curve is shown in fig.~7R of \cite{CMS:2016tvk}.\footnote{CMS also cuts on a $\Delta R_{opt}$ variable but they say this has the least discriminating power. We omit the cut on this variable for simplicity.} Our version of this overlaid on the CMS ROC curve is shown in fig.~\ref{fig:httv2}. We again see that the agreement is pretty good.

\section{Importance of the merge requirement}\label{Merge requirement}

Here we will elaborate further on the importance of the requirement ($\Delta R(t,q)<0.6$ in this paper, following \cite{CMS:2016tvk}) that the decay products of the top be ``fully merged". 
Tops failing the merge requirement generally result in fat jets that do not contain the full energy from the top quark decay. One can see this e.g.\ in fig.~2 of \cite{CMS:2016tvk} where histograms of the jet mass are shown with and without the merge requirement. Without the merge requirement, there is a clear peak and lower tail around the $W$ mass, indicating that some of the top jets are actually $W$ jets or the $b$ and only part of the $W$. 

\begin{figure}[t]
\begin{center}
\includegraphics[scale=0.5]{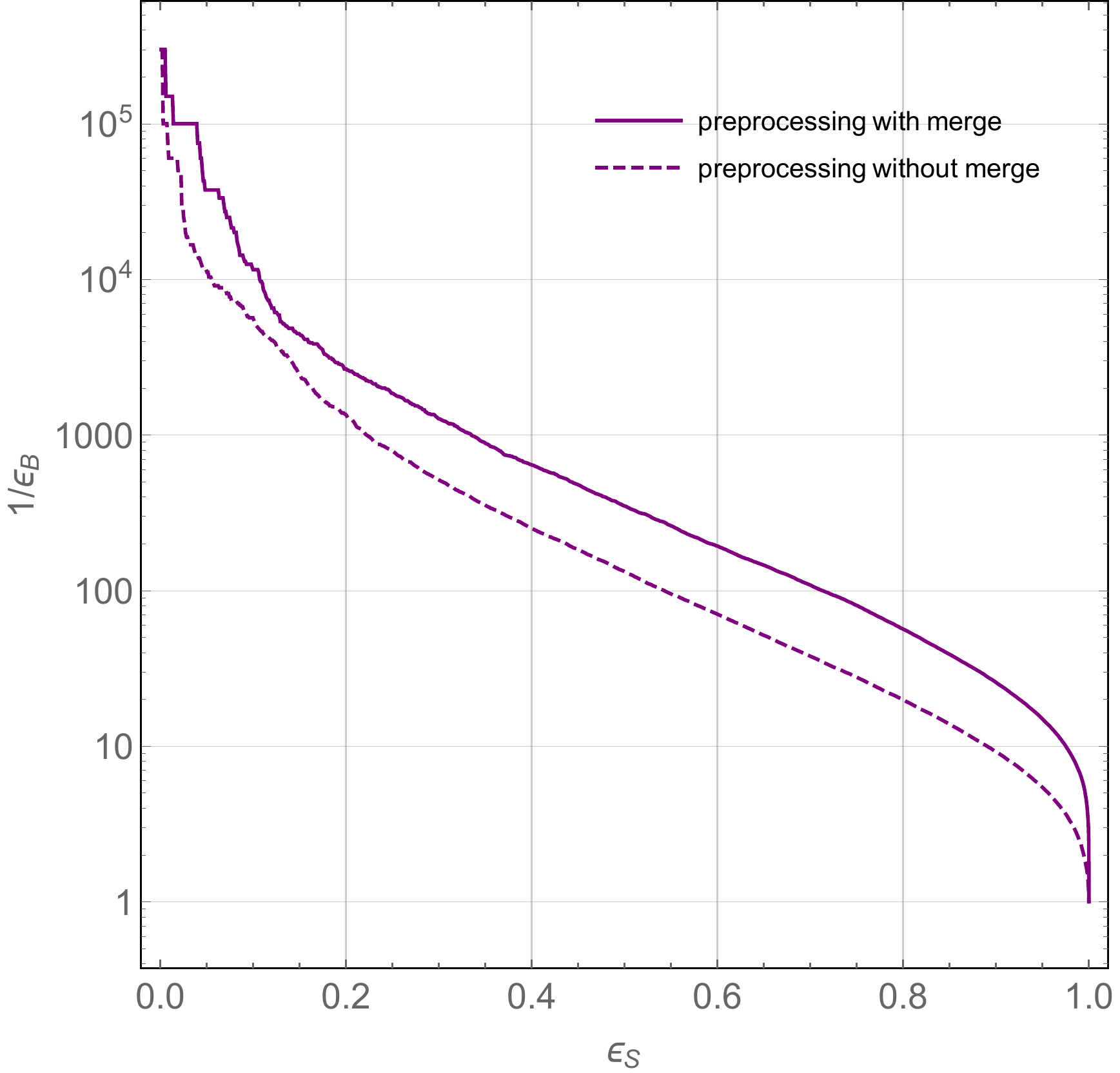}
\caption{ROC curves showing the performance of our top tagger on the CMS sample with and without the merge requirement.
}
\label{fig:mergereq}
\end{center}
\end{figure}

Restricting the signal sample to fully-merged tops will clearly boost the tagger performance, since the differences with QCD are more accentuated (the top jets are more top-like). This is illustrated in fig.~\ref{fig:mergereq} which compares the ROC curve for our CMS sample with preprocessing (the purple curve in fig.~\ref{fig:deeptopimprovements}) with and without the merge requirement. We see that the performance gain with the merge requirement is indeed substantial.

We remark in passing that the merge requirement could explain a 
puzzling discrepancy between the results in the DeepTop paper \cite{Kasieczka:2017nvn}  and the CMS note \cite{CMS:2016tvk}. Comparing the DeepTop ROC curve fig.~\ref{fig:deeptop} against the CMS ROC curves defined for a similar jet sample (fig.~7L of  \cite{CMS:2016tvk}), we see that the DeepTop tagger performs considerably worse, by a factor of $\sim 3$ or more. This is despite the DeepTop tagger being shown to outperform a BDT  trained on HTTV2 variables, which is among the best ROC curves shown in the CMS reference. We believe the crucial difference between the two ROC curves is the merge requirement. CMS requires their low $p_T$ tops to satisfy $\Delta R(t,q)<0.8$, while DeepTop \cite{Kasieczka:2017nvn} does not include this requirement.


\bibliographystyle{utphys}

\bibliography{jet_images_bib.bib}

\end{document}